\magnification=\magstep1
\hsize 6.0 true in
\vsize 9.0 true in
\voffset=-.5truein
\pretolerance=10000
\baselineskip=22truept

\font\tentworm=cmr10 scaled \magstep2
\font\tentwobf=cmbx10 scaled \magstep2

\font\tenonerm=cmr10 scaled \magstep1 
\font\tenonebf=cmbx10 scaled \magstep1

\font\eightrm=cmr8
\font\eightit=cmti8
\font\eightbf=cmbx8
\font\eightsl=cmsl8
\font\sevensy=cmsy7
\font\sevenm=cmmi7

\font\twelverm=cmr12  
\font\twelvebf=cmbx12
\def\subsection #1\par{\noindent {\bf #1} \noindent \rm}

\def\mid {\let\rm=\tenonerm \let\bf=\tenonebf \rm \bf}

\def\para{\par \vskip 12 pt}

\def\head{\let\rm=\tentworm \let\bf=\tentwobf \rm \bf}

\def\heading #1 #2\par{\centerline {\head #1} \smallskip
 \centerline {\head #2} \vskip .15 pt \rm}

\def\eight{\let\rm=\eightrm \let\it=\eightit \let\bf=\eightbf 
\let\sl=\eightsl \let\sy=\sevensy \let\m=\sevenm \rm}

\def\foots{\noindent \eight \baselineskip=10 true pt \noindent \rm}
\def\sexion{\let\rm=\twelverm \let\bf=\twelvebf \rm \bf}

\def\section #1 #2\par{\vskip 20 pt \noindent {\mid #1} \enspace {\mid #2} 
  \para \noindent \rm}

\def\abstract#1\par{\para \foots {\bf Abstract: \enspace}#1 \para}

\def\author#1\par{\centerline {#1} \vskip 0.1 true in \rm}

\def\abstract#1\par{\noindent {\bf Abstract: }#1 \vskip 0.5 true in \rm}

\def\sqr#1#2{{\vcenter{\vbox{\hrule height.#2pt
  \hbox {\vrule width.#2pt height#1pt \kern#1pt
  \vrule width.#2pt}
  \hrule height.#2pt}}}}

\def\n{\noindent}
\def\s{\smallskip}
\def\m{\medskip}
\def\b{\bigskip}
\def\c{\centerline}

\def\gne #1 #2{\ \vphantom{S}^{\raise-0.5pt\hbox{$\scriptstyle #1$}}_
{\raise0.5pt \hbox{$\scriptstyle #2$}}}

\def\ooo #1 #2{\vphantom{S}^{\raise-0.5pt\hbox{$\scriptstyle #1$}}_
{\raise0.5pt \hbox{$\scriptstyle #2$}}}


\line{\hfill IUCAA - 17/96}

\line{\hfill May 1996}
\m
\m

\c{\bf\mid  A  Conformal  Mapping  and  Isothermal  Perfect  Fluid  Model}
\b
\b
\b
\b
\b
\b
\b
\c{\bf Naresh Dadhich\footnote{$^* $}{E-mail : naresh@iucaa.ernet.in}}
\c{\bf Inter University  Centre for Astronomy \& Astrophysics}
\c{\bf P.O. Box 4, Pune-411007, India}
\b
\b
\b
\b
\b
\b
\b
                         
\c{\bf   Abstract }
\s 
\baselineskip=13truept 
Instead  of  conformal to  flat  spacetime, we  take  the  metric 
conformal  to a spacetime which can be thought of as ``minimally''
curved in the sense that free particles  experience  no 
gravitational  force  yet  it  has  non-zero  curvature. The  base 
spacetime  can  be written in the Kerr-Schild form  in  spherical 
polar  coordinates. The  conformal metric then  admits  the 
unique three parameter family of
perfect fluid solution which is static and  inhomogeneous. 
The density and pressure fall off in the curvature radial coordinates
as $R^{-2}, $  for unbounded
cosmological model with a barotropic equation of state. This is
the characteristic of isothermal fluid. We thus have an ansatz for
isothermal perfect fluid model. The solution
can also represent bounded fluid spheres.
  
\b
\b
\b
\b
\n PACS numbers: 0420, 9880

\vfill\eject
\baselineskip=22truept

\item{\bf 1.}{\bf Introduction}
\s
\n It  is well-known [1] that the Kerr-Schild form of  the  metric 
plays  an  important role in finding  vacuum, pure  radiation  and 
Einstein-Maxwell solutions in general relativity(GR). It turns out 
that  unless  the Kerr-Schild ansatz is generalised  the  perfect 
fluid  solutions cannot be found[2]. Senovilla and  coworkers[3-5] 
have obtained perfect fluid solutions by taking conformally  flat 
metric  as the seed metric in place of the original  flat  metric 
and thus generalising the KS ansatz. It reads as 

$$ \overline g_{ij} = g_{ij} + 2 H l_i l_j \eqno (1.1) $$

\n where $g_{ij} $  is the seed metric, $H $ is a scalar field and  
$l^i $ is  a  null 
vector  relative  to both $\overline g_{ij} $  and $g_{ij} $. 
When $g_{ij} = \eta_{ij} $, we  have  the 
original  KS  form. No perfect fluid solution can be  admitted  so 
long  as the seed metric is vacuum or its energy momentum  tensor 
has $l^i $ as an eigenvector[2].
\s

\n For  all  the perfect fluid solutions for the form  (1.1), the  seed 
metric is taken as conformally flat and $l^i $ as geodetic. That  means 
the  solution  can  be written as conformal to a  metric  in  the 
original  KS  form, which  may or may not be a solution of the 
Einstein  equation. We  shall  here consider  a  metric  which  is 
conformal  to  a  KS metric. The original metric  need  not  be  a 
solution  of the Einstein equation but has to be chosen  on  some 
physical   and  geometric  considerations. All  conformally   flat 
spacetimes will have $g_{ij} = e^{2U}~ \eta_{ij},~ U = U (x^i) $ 
and $H=0 $ in (1.1). There are  well-
known  conformally  flat perfect fluid  solutions ; e.g.  Friedman-
Robertson-Walker(FRW) and the Schwarzschild interior spacetimes.
\s

\n The  question  is what should be the metric in place of  flat  to 
conform to? What we wish to consider is,

$$ \overline g_{ij} = e^{2 U} (\eta_{ij} + 2 H l_i l_j),~ U = U(x^i) 
\eqno (1.2) $$

\n in which $H=0 $ gives conformally flat. The various forms of $H $
 will, for  the  base spacetime, give  vacuum, Einstein-Maxwell  and  pure 
radiation  solutions. What happens when $H $ is a  constant $\not= 0 $? Does 
the  original  metric now become flat? No, it  doesn't. In  fact  it 
represents an interesting situation which is free of the Newtonian 
gravity (free particles experience no acceleration), yet  spacetime 
is  curved  and  hence  it can  be  thought  of  as  ``minimally'' 
curved [6,7]. For vacuum solution $H $ satisfies the Laplace  equation 
corresponding to $R^0_0 = 0 $ (note that $R^0_0 = 0 $ is supposed to be
the analogue of the Newtonian equation $\bigtriangledown^2 \phi = 0 $)
and hence $H = const. \not= 0 $ will present so to say
a spacetime arising out of a  constant ``gravitational potential''.
This is why it is  free  of 
the  usual  gravitational force and hence very close (``minimally''
curved) to flat spacetime. How about  considering  a  metric 
conformal  to  it and enquire whether it is compitable  with  the 
perfect  fluid conditions? This is exactly what we wish to  do  in 
this paper.
\s

\n For  the  metric  (1.2) with $H=const. \not= 0 $, we shall  find  
the  most 
general  spherically symmetric perfect fluid solution which  will 
turn  out  to  be  static. It may be noted that when $H = 0 $,
 non-static perfect  fluid  solution like FRW is admitted but for $H $ 
different from  zero, 
even  when constant, only static solution is admitted. That means
$H \not= 0 $ implies severe constraints on the fluid solutions.
The  solution is  uniquely given by a three parameter family which   
can represent both bounded as well as unbounded distributions.
In the case of unbounded cosmological model density and pressure
satisfy a barotropic equation of state and fall off as inverse square
of the (curvature) radius. This is the characteristic behaviour for
an isothermal fluid model. 
\s
\n In section 2, we derive the metric and set up the field equations 
in  section  3  and find the general  perfect  fluid  solution. We
give some examples of fluid models in section 4. We 
conclude with a discussion.
\b

\item{\bf 2.} {\bf The metric}
\s

\n Consider the metric
$$ ds^2 = - dt^2 + dr^2 + r^2 d \Omega^2 + 2H (dt + dr)^2,
~ d \Omega^2 = d \theta^2 + sin^2 \theta ~d \varphi^2 \eqno (2.1) $$

\n which  is  in the KS form and we take $H=const \not= 0 $. 
This  metric can be transformed to the form

$$ ds^2 = -dt^2 + K^2 dr^2 + r^2 d \Omega^2,~ K^2 = (1 + 2H)^2/
(1 - 2H). \eqno (2.2) $$

\n The coordinate $t $ has been redefined to absorb the constant
coefficient. Recall  that  for  the vacuum solution $H $ satisfies  
the  Laplace 
equation for $R^0_0 = 0 $ and hence behaves like a gravitational 
 potential. It is not that $H $ is a solution of any Laplace equation
but is the solution of $R^0_0 = 0 $ that can be written as the
Laplace equation. It is this equation which is the analogue of 
the Newtonian equation defining potential. Hence $H $ may be
considered as gravitational potential.
 We  are taking $H $ constant  which  should  be 
 equivalent to constant 
potential  and  hence spacetime should be free of  the  Newtonian 
gravity  as can be verified by considering the geodesics  of  the 
above  metric. However the curvature is non-zero (it has  only  one 
curvature, $R^{23}_{~~23} = -(K^2-1)/K^2 r^2 $, which  is  an 
 invariant  of   spherical 
symmetry[7]) which  will  make  its presence felt  only  in  geosesic 
deviation  (tidal force) alone. Note that we have just  taken  the 
constant potential solution of the Laplace equation $R^0_0 = 0 $
and that  has 
generated  a curved spacetime and hence we would like to term  it 
as ``minimally'' curved[6,7]. Physically this can be thought of as 
the closest curved spacetime to flat spacetime.
\s

\n We shall consider the metric conformal to (2.2) and write it as

$$ ds^2 = e^{2U} (- dt^2 + K^2 dr^2 + r^2 d \Omega^2),~ U = U (r,t).
\eqno (2.3) $$

\n Note that both the Ricci and the curvature tensors for the metric 
(2.2)  have the only one non-zero component and they are  given  by 
[7],

$$ R^2_2 = {K^2 - 1 \over K^2 r^2} = - R^{23}_{~~23}. \eqno (2.4) $$

\n Let us write from [1], 

$$ \overline R_{ab} = R_{ab} - 2 Y_{ab} - g_{ab} Y^c_c \eqno (2.5) $$                                            
$$ Y_{ab} = U_{;ab} - U_{,a} U_{,b} + {1 \over 2} g_{ab}
U_{, c} U^{,c} \eqno (2.6) $$  

\n where $R_{ab} $  and $\overline R_{ab} $ refer to $g_{ij} $   and 
$\overline g_{ij} = e^{2 U} g_{ij} $. Then  for  the 
metric (2.3) we have 

$$ e^{2U} ~\overline R^0_0 = - {1 \over K^2} \bigg(U^{\prime \prime} + 
2 U^{\prime 2} + {2 U^{\prime} \over r} \bigg) + 3 \ddot U 
~~~~~~~~~~~~~~ \eqno (2.7) $$

$$ e^{2U} ~\overline R^1_1 = - {1 \over K^2} \bigg(3 U^{\prime \prime} + 
 {2 U^{\prime} \over r} \bigg) + \ddot U + 2 \dot U^2
~~~~~~~~~~~~~~~~~~~~ \eqno (2.8) $$

$$ e^{2U}~ \overline R^2_2 = - {1 \over K^2} \bigg(U^{\prime \prime} + 
2 U^{\prime 2} + {4 U^{\prime} \over r} - {K^2 - 1 \over r^2} \bigg)
+ \ddot U + 2 \dot U^2 \eqno (2.9) $$

$$ \overline R_{01} = -2 (\dot U^{\prime} - \dot U U^{\prime}) \eqno (2.10) $$
 
\n where $\dot U = \partial U/ \partial t $ and $U^{\prime} = \partial U/
\partial r $.
 
\b

\item{\bf 3.} {\bf The general perfect fluid solution}
\s

\n The field equation for a perfect fluid distribution reads as

$$ R_{ik} = 8 \pi \bigg[ (\rho + p) u_i u_k + {1 \over 2}
(\rho - p) g_{ik} \bigg] \eqno (3.1) $$

\n which for a comoving velocity field $u = e^U dt $
implies $R_{01} = 0 $  and $R^1_1 = R^2_2 $ (dropping overhead
bar in (2.7) - (2.10)). From (2.10) we get

$$ \dot U^{\prime} - \dot U U^{\prime} = 0 $$

\n which integrates to give
$$e^{-U} = f(t) + g(r). \eqno (3.2) $$

\n Though we began with the KS form but we are ultimately considering
the metric (2.3) which is orthogonal and spherically symmetric. Hence
we are justified in employing the comoving coordinates.
Substituting  (3.2) in $R^1_1 = R^2_2, $  following from  (2.8)  
and (2.9), leads to

$$ \bigg(-g^{\prime \prime} + {g^{\prime} \over r} + 
{K^2 - 1 \over K^2 r^2} \bigg)
{K^2 r^2 \over K^2 - 1} = - f(t) \eqno (3.3) $$

\n which  clearly  implies  $f(t) = const $. This  means  the  conformal 
function $ U $ must only be a function of $r $ alone and the  spacetime 
then  becomes static. Note that this conclusion follows from (3.3) 
only  when  $K \not= 1, $ and when $K = 1, $ (3.3)  does  admit  non-static 
solutions as for FRW. This precisely demostrates the role  played 
by the non-flat character (which is introduced through $K \not= 1 $)   
 of the original metric.
\s
\n With $U = U(r) $, the only differential equation to be solved is

$$ U^{\prime \prime} - U^{\prime 2} - {U^{\prime} \over r}
+ {K^2 - 1 \over 2r^2} = 0 \eqno (3.4) $$

\n which admits the general solution

$$ e^{-U} = c_1 r^n + c_2 r^{2-n},~K^2 = 1 + 2n (n-2) \eqno (3.5) $$

$$ 8 \pi \rho K^2 e^{2U} r^2 = 2 {(2n - 1) n c_1 + (2 - n)(3 - 2n) c_2 \mu
\over c_1 + c_2 \mu} - 3 \bigg({n c_1 + (2-n) c_2 \mu \over c_1 + 
c_2 \mu} \bigg)^2 
\eqno (3.6) $$

$$ 8 \pi p K^2 e^{2U} r^2 ~=~ -2 ~{n^2 c_1 ~+~ (2-n)^2 c_2 
\mu \over c_1 ~+~ c_2 \mu}
~+~ 3 \bigg({ n c_1 + (2-n) c_2 \mu \over c_1 ~+~ c_2 \mu} \bigg)^2~~~~~~~~~~~ \eqno (3.7) $$

\n where $\mu = r^{2(1-n)} $.
\s
\n From (3.5), the window $1 - 1/\sqrt 2 <
n < 1 + 1/\sqrt 2 $ is prohibited for $n $. $K = 1 $ for $n = 0,
2 $ and then the metric turns conformally flat. $\rho $ will always remain 
positive for $c_1/c_2 > 0 $ and
 $n < 0 $ or $n> 2 $. It may be noted from (3.5) that we need
to consider either $n < 0 $ or $n > 2 $ because one can be
transformed into the other simply by $c_1 \longleftrightarrow c_2 $.
>From (3.6) and (3.7) it can be seen that the fluid
cannot admit a barotropic equation of state, $\rho = kp $ unless 
either $c_1 $ or $c_2 $ vanish. In that case the solution is cosmological
for $p $ cannot be made zero for a finite $r $. 
\s
Let us examine the condition for a bounded sphere, i.e. when 
$p = 0 $ which implies 

$$ (2-n)^2 \mu^2 + 2 {c_1 \over c_2} \bigg(2(2-n) (2n-1) - n^2 \bigg) \mu
+ n^2 {c_1^2 \over c^2_2 } = 0. \eqno (3.8) $$

\n It will admit a real solution only if $3n^2 - 6n + 2 \geq 0 $
which rules out the window $1 - 1/\sqrt 3 \leq n \leq 1 + 1/\sqrt 3 $
for $n $. This range is entirely covered by the earlier forbidden
range $1 - 1/\sqrt 2 < n < 1 + 1/\sqrt 2 $. Hence so long as $c_1 $
and $c_2 $ are non-zero $p $ will always vanish at some finite $r $
defining the boundary of fluid sphere. It is obvious that when one of
$c_1 $ and $c_2 $ is zero, $p $ cannot vanish at any $r $ and hence the fluid
will have no boundary. Thus we have a bounded fluid sphere for
$c_1 $ and $c_2 \not= 0 $ and unbounded for one of them being zero.
\s
\item{\bf 4.}{\bf Fluid models }
\s
\item{\it 4.1} {\it Unbounded and cosmological :} 
\s
In here we
have to take one of $c_1 $ and $c_2 $ zero and then (3.5) will represent
the only one solution with $n $ or $(2 - n )$ as the free parameter.
\s
\n Without any loss of generality let us set $c_1 = 1 $ and
$c_2 = 0 $ and have

$$ ds^2 = r^{-2n} (- dt^2 + K^2 dr^2 + r^2 d \Omega^2) \eqno (4.1) $$

\n and

$$ 8 \pi \rho = {n^2 - 2n \over K^2 r^{2(1-n)}},~ 8 \pi p = {n^2 \over K^2
 r^{2(1-n)}},
K^2 = 1 + 2n (n - 2). \eqno (4.2) $$

\n Note that physically relevant
radial coordinate is the one that occurs as $R^2 $ coefficient of
$d \Omega^2 $ in the metric; i.e. $R = r^{1-n} $. Thus
in terms of the curvature radial coordinate density and pressure fall 
off as $R^{-2} $ (This $R $ should not be confused with
the scalar curvature $R $). The inverse square fall off is characteristic
of isothermal equilibrium [8] of fluid with a barotropic equation
of static.
\s
\n The positivity of $\rho $ is ensured by $n > 2 $ or $n < 0 $
which is permissible for it lies outsided the prohibited window.
The equation of state for fluid is given by

$$ \rho = {n - 2 \over n}~ p \eqno (4.3) $$

\n Now $\rho \geq 3p $ requires $n \leq -1 $ which means $k^2
\geq 7 $. The spacetime becomes flat for $n = 0 $. It may be
noted that $\rho \geq 3p $ automatically ensures $\rho \geq 0 $
(i.e. $n \leq -1) $.
\s
\n Though $\rho $ diverges
as $R \longrightarrow 0 $ but total mass contained inside a radius $R $
remains always finite and goes to zero with $R $.
Also note that $dp/d \rho $ is always less than $1 $, indicating velocity
of light is always greater than velocity of sound. 
\s
\item{\it 4.2}{\it Fluid spheres}
\s
\n For the bounded fluid distributions, $c_1 $ and $c_2 $ must be non-zero.
For positivity of density $c_1/c_2 > 0 $ and $n \leq 0 $ or $n \geq 2 $,
where the equality implies $K = 1 $ and will make the spacetime
conformally flat. For both $n = 0 $ and $n = 2 $, we get the same
solution with the role of $c_1 $ and $c_2 $ interchanged.
\s
\n When $n = 0, ~K = 1, $ we have

$$ e^{- U} = c_1 + c_2 r^2, \eqno (4.4) $$

$$ 2 \pi \rho = 3 c_1 c_2  \eqno (4.5) $$

$$ 2 \pi p = c_2 (-2c_1 + c_2 r^2)  \eqno (4.6) $$

\n which means $c_2 r^2 \geq 2 c_1 $. This will imply an upper (lower)
bound on $r $ depending upon both $c_1, c_2 < 0 (> 0 ) $. The boundary
of the sphere is defined by $r^2_0 = 2 c_1/c_2 $. The solution is
conformally flat because $K = 1 $. It is a uniform density conformally
flat fluid sphere. It is therefore a transform of the Schwarzschild interior
solution.
\s
\n Let us consider an example of conformally non-flat fluid sphere. Take
$n = -1 $, say, then $K^2 = 7 $, and

$$ e^{- U} = c_1 r^{-1} + c_2 r^3 \eqno (4.7) $$

$$ 8 \pi \rho = {3 \over 7 } (c^2_2 r^4 + 18 c_1 c_2 + c^2_1
r^{-4} ) \eqno (4.8) $$

$$ 8 \pi p = {1 \over 7 } (9 c^2_2 r^4 - 38 c_1 c_2 + c^2_1 r^{-4})
 \eqno (4.9) $$

\n The fluid boundary is defined by $r^4_0 = (19 \pm \sqrt 10) c_1
/ 9 c_2 $. This means either $r^4 \leq \bigg({19 - \sqrt 10 \over 9} \bigg)
{c_1 \over c_2} $ or $r^4 \geq \bigg({19 + \sqrt 10 \over 9} \bigg)
{c_1 \over c_2} $. But $\rho \geq p $ requires $r^4 \leq \bigg({46 \over
3}  \bigg) {c_1 \over c_2}$ which rules out the latter case. There 
are two free parameters $c_1, c_2 $ in the solution and hence
it could easily be matched to the Schwarzschild exterior solution
at $r^4_0 = \bigg({19 - \sqrt 10 \over 9} \bigg) {c_1 \over c_2} $. Thus
this solution can describe the interior of a star in hydrostatic
equilibrium. As in the cosmological case $\rho $ and $p $ diverge
as $r \longrightarrow 0 $ but the mass contained in a radius $r $ 
remains finite and goes to zero with $r $. 
\b
\item{\bf 5.} {\bf Discussion}
\s
\n The  solution  obtained above is of the Tolman type [9]  but  the 
important point is that it follows as the unique solution of  the 
metric  ansatz (2.3); i.e. the metric conformal to a KS  spherically 
symmetric   metric   free  of  the  Newtonian   gravity. There are
well-known conformally flat perfect fluid solutions. Here we retain the 
conformal character but let go flatness of the original metric and yet 
essentially keeping it free of usual gravity. It exhibits in the context
of perfect fluid spacetimes what happens 
when flatness of the base metric is relaxed yet retaining the essential
physical character of spacetime. It is interesting that it leads  to 
the unique solution representing a three parameter $(n, c_1 $
and $c_2 $) family.
\s
\n The  perfect fluid is in hydrostatic equilibrium  resulting  from 
pressure gradient balancing radial acceleration. The general solution
can provide both the bounded fluid spheres as well as unbounded
cosmological distribution depending upon the choice of free
parameters. 
\s
\n The remarkable feature of this family is that in the cosmological
case it gives $\rho $ and $p $ satisfying a barotropic equation of
state and falling off inverse square of the curvature radial
coordinate which is the physically relevant radius. This behaviour is
characteristic of an isothermal fluid sphere [8]. Recently it has been
argued [10] that the ultimate end state of the Einstein-deSitter model
would approximate to an isothermal fluid sphere. As mentioned earlier,
in such models though density may diverge, but total mass contained
inside a finite radius will however be finite and will go to zero
as radius goes to zero.
\s
\n Finally we would like to say that the ansatz (2.3) is a  remarkably 
simple  way  of obtaining a static  inhomogeneous and isothermal
 perfect  fluid 
model with a barotrpic equation of state and density falling  off 
as  $R^{-2} $. It is remarkable that (2.3) admits the 
unique  solution with these specific properties. Our  ansatz may be
taken as characterising such a behaviour.
\b
\n{\bf Acknowledgement :} I would like to thank the unanimous
referee for pointing an oversight error which has led to 
stronger result than the one in the original version.

\b
\c{\bf References}
\b
\item{[1]} Kramer D, Stephani H, Herlt E and MacCallum M A H
1980 {\it Exact solutions of Einstein's field equations} (Cambridge:
Cambridge University Press)
\s
\item{[2]} Senovilla J M M and Sopuerta C F 1994 Class.Quantum Grav.  11 
     2073.
\s
\item{[3]} Martin J and Senovilla J M M 1986 {\it J. Math. Phys.}
{\bf 27} 265  
\s
\item{[4]} Senovilla J M M 1986 {\it Ph.D. Thesis} Universidad
de Salamanca
\s
\item{[5]} Martin-Pascual F and Senovilla J M M 1988 {\it J. Math. Phys.}
{\bf 29} 937
\s
\item{[6]} Dadhich N (see Gr-14  Abstracts, A.98) On the
Schwarzschild field and little beyond, to be published
\s
\item{[7]} Dadhich N 1970 {\it Ph.D. Thesis}, Poona University, unpublished
\s
\item{[8]} Chandrasekhar  S 1939 Introduction to the study  of  Stellar 
      Structure(Chicago:Univ. of Chicago)
\s
\item{[9]} Tolman R C 1939 Proc. Nat. Acad. Sci. 20 169
\s
\item{[10]} Saslaw W C, Maharaj S D and Dadhich N (1996) An isothermal
Universe, submitted to Ap.J.

\bye